\documentclass[fleqn,usenatbib]{mnras}

\usepackage{newtxtext,newtxmath}
\usepackage[T1]{fontenc}

\usepackage[export]{adjustbox}
\usepackage[utf8]{inputenc}
\usepackage{subcaption}
\usepackage{graphicx}	% Including figure files
\usepackage{amsmath}	% Advanced maths commands
\usepackage{amssymb}	% Extra maths symbols

\title[The \textit{TESS} light curve of 1SWASPJ0113+31]{The \textit{TESS} light curve of the eccentric eclipsing binary 1SWASP J011351.29+314909.7 -- no evidence for a very hot M-dwarf companion}

\author[M. I. Swayne et al.]{
Matthew I. Swayne,$^{1}$\thanks{E-mail: m.i.swayne@keele.ac.uk}
Pierre F. L. Maxted,$^{1}$
Vedad  Kunovac Hod\v{z}i\'{c}$^{2,3}$\thanks{Fulbright Fellow}
\newauthor and Amaury H. M. J. Triaud$^{2}$
\\
$^{1}$Astrophysics Group, Keele University, Staffordshire, ST5 5BG, UK\\
$^{2}$School of Physics and Astronomy, University of Birmingham, Edgbaston, Birmingham B15 2TT, UK\\
$^{3}$Department of Astronomy and Astrophysics, University of Chicago, 5640 S. Ellis Avenue, Chicago, IL 60637, USA
}

\date{Accepted 2020 July 2. Received 2020 June 10; in original form 2020 March 30}

\pubyear{2020}

\begin{document}
\label{firstpage}
\pagerange{\pageref{firstpage}--\pageref{lastpage}}
\maketitle

\begin{abstract}
A 2014 study of the eclipsing binary star 1SWASPJ011351.29+314909.7 (J0113+31) reported an unexpectedly high effective temperature for the M-dwarf companion to the 0.95-M$_{\odot}$ primary star.
The effective temperature inferred from the secondary eclipse depth was $\sim$600 K higher than the value predicted from stellar models.
Such an anomalous result questions our understanding of low-mass stars and might indicate a significant uncertainty when inferring properties of exoplanets orbiting them.
We seek to measure the effective temperature of the M-dwarf companion using the light curve of J0113+31 recently observed by the \textit{Transiting Exoplanet Survey Satellite} (\textit{TESS}).
We use the {\fontfamily{qcr}\selectfont pycheops} modelling software to fit a combined transit and eclipse model to the \textit{TESS} light curve.
To calculate the secondary effective temperature, we compare the best-fit eclipse depth to the predicted eclipse depths from theoretical stellar models.
We determined the effective temperature of the M dwarf to be ${\rm T}_{\rm eff,2}$ = 3208 $\pm$ 43 K, assuming $\log g_2$ = 5, [Fe/H] = $-0.4$ and no alpha-element enhancement.
Varying these assumptions changes ${\rm T}_{\rm eff,2}$ by less than 100 K.
These results do not support a large anomaly between observed and theoretical low-mass star temperatures. 
\end{abstract}


%%%%%%%%%%%%%%%%%%%%%%%%%%%%%%%%%%%%%%%%%%%%%%%%%%

%%%%%%%%%%%%%%%%% BODY OF PAPER %%%%%%%%%%%%%%%%%%

\begin{figure*}
    \centering
    \begin{subfigure}{0.49\textwidth}
        \centering
        \includegraphics[width=0.95\linewidth,height=0.95\linewidth]{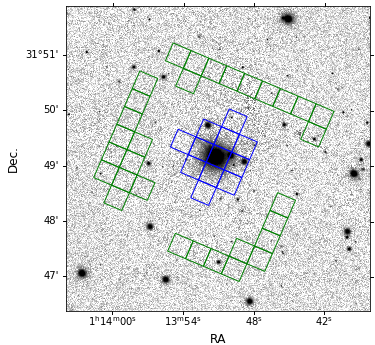} 
        \label{fig:subim1}
    \end{subfigure}\begin{subfigure}{0.49\textwidth}
        \centering
        \includegraphics[width=0.95\linewidth]{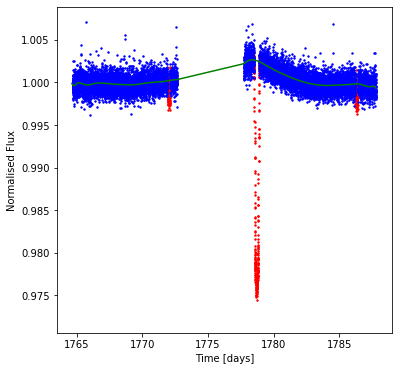}
        \label{fig:subim2}
    \end{subfigure}
     
    \caption{\textit{Left}: \textit{TESS} pixels for its observation of J0113+31 overlaid onto an image of the area around the object from the PanSTARSS image server, \citep{flewelling}.
    J0113+31 is the bright, central star, the TESS photometric aperture is in blue and the pixels used to calculate the background flux are in green.
    \textit{Right}: The TESS light curve of J0113+31. 
    The light curve is shown in blue with the eclipse and transit events masked in detrending shown in red. 
    The polynomial used to detrend the light curve is overlaid in green.}
    \label{fig:figure1}
\end{figure*}

\section{Introduction}

   One of the most important factors in correctly characterising an exoplanet is to understand its host star.
   The parameters of an orbiting exoplanet are, in most cases, inferred from its effect upon the signal of its stellar host, most commonly through the transit or radial velocity methods.
   The host star properties are most often obtained by matching observable star properties to stellar evolution models \citep[e.g.][]{Baraffe98,Dotter}.
   Thus, if these models are erroneous, and with them our understanding of the primary star, so too will any exoplanet observations that are inferred from them.
   This raises a possible issue regarding low-mass stars.
   Low-mass stars suffer from a lack of data compared to other brighter sources.
   Direct measurements of stellar mass and radius are uncommon and of temperature rarer still.
   As low mass stars are being looked upon more and more as favourable targets for exoplanet detection and characterisation \citep{charbonneau2007dynamics, quirrenbach2014carmenes,delrez} this could be a great problem for both current and future observations.
   Recently, the EBLM Project \citep{triaud} has been launched to start to address this problem.
   Its aim is to characterise around 200 low-mass eclipsing binary (EBLM) systems discovered in the SuperWASP survey to better understand M dwarf stars. 
   
   One study in the EBLM project, \cite[][GMC+2014 hereafter]{Yilen}, has reported derivations of the mass, radius and temperature of the eclipsing M-dwarf system 1SWASPJ011351.29+314909.7 (J0113+31 hereafter).
   They inferred a much higher M dwarf temperature than predicted by theoretical models.
   A similar issue was noted by \cite{Ofir} in their analysis of KIC 1571511B.
   If this inconsistency is a wider trend it could result in the incorrect characterisation of exoplanets in low-mass star systems.
   J0113+31 was recently observed by the \textit{TESS} mission \citep{Ricker}.
   This allows us to see if we can reproduce this anomalous secondary temperature measurement.
   In this Letter we present the analysis of the \textit{TESS} light curve of J0113+31.
   After fitting the observed light curve using Monte Carlo Markov Chain (MCMC) techniques, we then compared the observed secondary eclipse depth to those predicted by theoretical stellar spectra.
   We find that our observed secondary effective temperature does not agree with the unexpectedly high temperatures seen in GMC+2014, implying a value expected for a low-mass M dwarf.
   
\section{Observation}
   The \textit{TESS} survey is split into 26 overlapping $90^{\circ} \times 24^{\circ}$ degree sky sectors over both northern and southern hemispheres, with each observed for approximately one month.
   The eclipsing binary J0113+31 (TIC 400048097) was observed in Sector 17 of the survey as part of the Guest Investigator programs G022039 and G022062, with 2-minute cadence data made available.
   J0113+31 is a bright (\textit{V} = 10.1) eclipsing binary star composed of a G0-2 V, metal-poor (${\rm [Fe/H]}=-0.4$) primary star and a much fainter M-dwarf companion with a mass of about 0.2 M$_{\odot}$. 
   The orbital period is approximately 14.3 days and the orbit is eccentric ($e\approx 0.3$).
   We downloaded the light curve from the Mikulski Archive for Space Telescopes (MAST)\footnote{\url{https://mast.stsci.edu}} web service.
   We used the PDCSAP flux data for our analysis.
   Any cadences in the light curve with severe quality issues were ignored using the "default" bitmask 175 \citep{tenenbaum}.
   We downloaded the target pixel file for the target and overlaid the \textit{TESS} aperture used onto a map of the local sky area in order to confirm that the Science Processing Operations Center (SPOC) pipeline accounted for the presence of any contaminating stars.
   From Figure \ref{fig:figure1} it can be seen that there are 3 faint stars within the photometric aperture.
   The flux from J0113+31 relative to the total flux of all stars in the photometric calculated from the TESS magnitudes from the TESS input catalogue \citep{stassun2019revised} is 0.9722.
   This is similar to the reported crowding metric used for J0113+31 of 0.9695 so we are satisfied that the PDCSAP flux had been corrected for this contaminating flux.
   In addition, we observed a slight stellar variation in the light curve.
   We removed the resultant low-frequency noise by masking the transits events, fitting a polynomial of order 25 and dividing the unmasked light curve by the resulting function, shown in Figure \ref{fig:figure1}.

\section{Analysis/Results}
    To create the models needed for light curve fitting we used {\fontfamily{qcr}\selectfont pycheops}\footnote{\url{https://pypi.org/project/pycheops/}}, a python module developed for analysis of data from the \textit{CHEOPS} mission \citep{Cessa}.
    The transit model uses the qpower2 algorithm \citep{maxted2019q} to calculate the transit light curve assuming a power-2 limb darkening law.
    The parameters used in the model are: the time of mid-primary eclipse $T_0$, the transit depth $D = k^2 = R_2^2/R_1^2$ where $R_2$ and $R_1$ are the radii of the secondary and primary stars, the impact parameter $b = a \cos{i}/R_1$ where $i$ is the orbital inclination and $a$ is the semimajor axis, the transit width $W = \sqrt{(1+k)^2 - b^2} R_1/(\pi a)$, the eccentricity and argument of periastron dependent parameters $f_s = \sqrt{e} \sin{(\omega)}$ and $f_c = \sqrt{e} \cos{(\omega)}$, the eclipse depth $L$ and the limb-darkening parameters $h_1$ and $h_2$ as defined by \cite{maxted2018}.
    The light curve only includes one primary and two secondary eclipses so we fixed the orbital period at the value $P$ = 14.2769001 d from GMC+2014.
    As $h_2$ did not converge to a value during the MCMC fit, we fixed it at a value obtained by an interpolator in-built in {\fontfamily{qcr}\selectfont pycheops}.
    This interpolates a value of $h_2$ from a data table presented in \cite{maxted2018} based on the limb-darkening profiles from the STAGGER-grid \citep{magic2015stagger}.
\begin{figure}
    \includegraphics[width=0.95\linewidth,height=\linewidth]{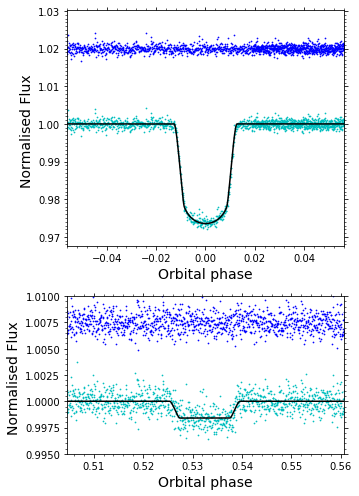} 
    \caption{Fitted normalised light curve of J0113+31 in phase intervals around the transit and eclipse events.
    In both plots the observed light curve is displayed in cyan, the best fit model is shown in black and the residual of the fit is presented in blue.}
    \label{fig:figure2}
\end{figure}
\begin{figure*}
    \centering
    \begin{subfigure}{0.49\textwidth}
        \centering
        \includegraphics[width=0.95\linewidth,height=0.95\linewidth]{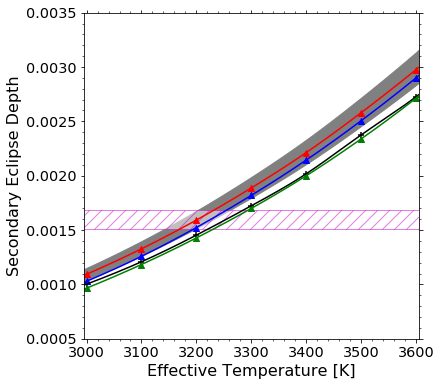} 
        \label{fig:subim3}
    \end{subfigure}\begin{subfigure}{0.49\textwidth}
        \centering
        \includegraphics[width=0.95\linewidth,height=0.95\linewidth]{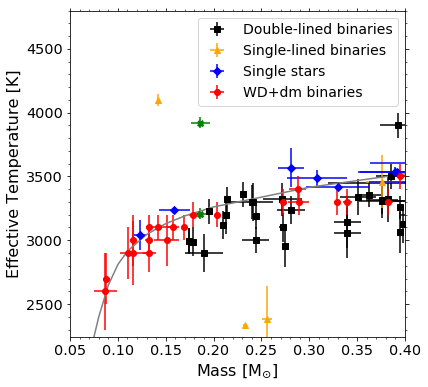}
        \label{fig:subim4}
    \end{subfigure}
    \caption{\textit{Left}: The secondary eclipse depths predicted using the PHOENIX (\citet{Husser}, triangles) and BT-Settl-CIFIST (\citet{Baraffe15}, crosses) theoretical stellar spectra. 
    All models assume ${\rm T}_{\rm eff,1}$ = 6000 K, $\log g_1$ = 4.00 dex, $\log g_2$ = 5.00 dex, and no alpha element enhancement. 
    We varied the metallicity between these sets with [Fe/H] = $-0.5$, 0.0 and 0.5 dex for red, blue and green markers respectively for PHOENIX models. 
    The grey area represents a 100 K uncertainty in ${\rm T}_{\rm eff,1}$. 
    The magenta line shows the fitted eclipse depth from the \textit{TESS} light curve, L = 0.00160 $\pm$  0.00009. 
    \textit{Right}: A cutout of the stellar mass versus effective temperature diagram from \citet{parsons2018scatter}, with our result and the result from GMC+2014 highlighted (green crosses). 
    The type of system is displayed by different colours and symbols. 
    The theoretical relation from \citet{Baraffe15} for an age of 1 Gyr is plotted in gray.}
    \label{fig:figure3}
\end{figure*}
\begin{table*}
    \centering
      \caption{The reported orbital parameters from \citet{Maxted}, GMC+2014 and the parameters calculated by our {\fontfamily{qcr}\selectfont pycheops} and {\fontfamily{qcr}\selectfont ellc} fits.}
         \label{Params}
    $$ 
         \begin{array}{lrrrr}
            \hline
            \noalign{\smallskip}
            \text{}      & \multicolumn{1}{l}{\text{GMC+2014}} & \multicolumn{1}{l}{\text{Maxted (2016)}} & \multicolumn{1}{l}{\text{{\fontfamily{qcr}\selectfont pycheops} fit}} & 
            \multicolumn{1}{l}{\text{{\fontfamily{qcr}\selectfont ellc} fit}}\\
            \noalign{\smallskip}
            \hline
            \noalign{\smallskip}
            R_1/a & 0.0534 \pm 0.0021 & 0.0533 \pm 0.0004 & 0.0540 \pm 0.0010 & 0.0536 \pm 0.0006      \\
            R_2/a & 0.0081 \pm 0.0004 & 0.00783 \pm 0.00008 & 0.0083 \pm 0.0002 & 0.0082 \pm 0.0001   \\
            i \text{ ($^{\circ}$)} & 89.084 \pm 0.037 & 89.09 \pm 0.05 & 88.980 \pm 0.103 & 89.062 \pm 0.064 \\
            L_{J} & 0.00737 \pm 0.00024 & 0.00749 \pm 0.00018 & -- & --    \\
            L_{\textit{TESS}} & -- & -- & 0.00160 \pm 0.00009 & 0.00164 \pm 0.00006             \\
            e & 0.3098 \pm 0.0005  & 0.3096 \pm 0.0007 & 0.3138 \pm 0.0151 & 0.3090 \pm 0.0090 \\
            \omega \text{ ($^{\circ}$)} & 278.85 \pm 1.29 & 278.9 \pm 0.03 & 278.88 \pm 0.47 & 279.01 \pm 0.30\\
            \noalign{\smallskip}
            \hline
         \end{array}
    $$
\end{table*}
    We used the python module {\fontfamily{qcr}\selectfont emcee} \citep{Foreman} to sample the posterior probability distribution of our model parameters.
    We sampled a chain of 480 walkers each going through 6000 steps, starting at values determined by a least-squares fit and with step-sizes set to suitable values for each parameter.
    To allow the walkers to settle into the probability distributions we performed a burn-in of 500 steps before the sampling.
    To ensure adequate sampling was performed the number of steps chosen was $\sim$65-75 times longer than the autocorrelation length of each fitted parameter chain.
    To ensure independent random samples from their posterior probability distributions, each parameter chain was thinned by half the minimum parameter autocorrelation length.
    The parameter values given in Table~\ref{Params} are the mean and standard deviation of each of the thinned model parameter chains.
    The light curve fit and residuals for these parameter values are shown in Figure \ref{fig:figure2}.
    We verified our analysis by performing an independent fit using the eclipsing binary light curve model, {\fontfamily{qcr}\selectfont ellc} \citep{Maxted}, as implemented in a package called {\fontfamily{qcr}\selectfont amelie} \citep[e.g.][]{hodvzic,triaud2020eclipsing}.
    We find fully consistent results between using the two light curve models as shown in Table \ref{Params}.
    
\section{Discussion}
    To convert the parameters from our light curve model to an estimate of  ${\rm T}_{\rm eff}$ for the M-dwarf star we used a similar method to GMC+2014.
    This involved comparing the observed secondary eclipse depth with the expected depth determined using PHOENIX model atmospheres \citep{Husser}.
    In brief, we integrate the flux of the primary star over the \textit{TESS} bandpass and proceed to calculate fluxes for the secondary over a range of different temperatures using the same technique.
    We assume [Fe/H] = $-0.5$ dex, ${\rm T}_{\rm eff,1}$ = 6000 K, $\log g_1$ = 4.00 dex, $\log g_2$ = 5.00 dex, (as in GMC+2014), and no alpha element enhancement.
    The predicted eclipse depth is then \mbox{$\Delta_2 = D\frac{F_2}{F_1}$}, where $F_1$ and $F_2$ are the integrated fluxes for the primary and secondary stars.
    Using this method, eclipse depths were determined for ${\rm T}_{\rm eff,2}$ values from 2500 to 4000 K.
    For further comparison showing the effect of different metallicities, predicted eclipse depths were also calculated using [Fe/H] = 0.0 dex and 0.5 dex.
    As shown in Figure \ref{fig:figure3}, the eclipse depth predicted by the theoretical stellar models would indicate an effective temperature far lower than that found by GMC+2014 for all three cases we calculated, with no difference in metallicity enough to reconcile our results with their derived temperature of 3922 K.
    To provide a further comparison we also calculated eclipse depths using BT-Settl-CIFIST model spectra, comparing it with those obtained by PHOENIX using a consistent [Fe/H] = 0.
    Again, the observed difference is not enough to account for the anomalous temperatures seen in GMC+2014.
   
   For our best estimate of the M-dwarf effective temperature we decided to use the value of [Fe/H] = $-0.4$ $\pm$ 0.04 provided in GMC+2014, obtaining it through linear interpolation of eclipse depths at different metallicities using the PHOENIX derived values.
   Due to the uncertainty in abundances when varying stellar parameters \citep{jofre}, we increased the [Fe/H] error to $\pm$0.1 dex.
   We calculated the uncertainty in ${\rm T}_{\rm eff,2}$ by combining uncertainties in depth, ${\rm T}_{\rm eff,1}$ and metallicity.
   Adding these uncertainties in quadrature we obtained a final effective temperature, ${\rm T}_{\rm eff,2}$ = 3208 $\pm$ 43K.
   As shown in Figure \ref{fig:figure3}, this is the effective temperature expected for this star given its mass.
   
   As the result in GMC+2014 had been so unexpected they had discussed and discounted several sources of potential theoretical error, either being not feasible or not having enough of an effect to cause a temperature $\sim$600K warmer than expected.
   Therefore, to examine the possible causes for this inconsistency with our results, we first verified them using an independent code ({\fontfamily{qcr}\selectfont ellc}).
   We then looked for any problems in our own integration of the theoretical models.
   We did this by reproducing our eclipse depth predictions but integrating in the same bandpass as that used by GMC+2014, specifically that of the FLAMINGOS instrument used to observe their secondary eclipse data.
   This correctly reproduces their theoretical expected depths, ruling out problems in this element of our analysis.
   We also tested our method for dividing out variation in the light curve by observing whether the method is sensitive to the order of the polynomial used in removing slow flux variations in the light curve.
   If we use a polynomial of order 10 instead of 25 we find that the value of ${\rm T}_{\rm eff,2}$ changes by only 3K, i.e., not enough to put our overall conclusion in doubt.
   
  We then searched for inconsistencies in the observational measurements of the two studies.
  One contributing factor could be the issue of metallicity and how it effects observations at different wavelength regimes.
  At a fixed mass, a metal-rich star is predicted to see a decrease in luminosity caused by the increased opacity.
  However this increase in opacity does not necessarily lead to a reduction in flux in all bands.
  \cite{Mann} finds that in the \textit{K} band this trend could be weakened or reversed due to the increased opacities occurring in the visible rather than the near-infrared, causing a larger amount of the flux to escape.
  They display the flux ratio of metal-poor and metal-rich stars in different wavelength regimes, finding a change from 1.2 to 1.0 between \textit{r$^{\prime}$} and \textit{K} bands.
  As the \textit{TESS} satellite operates from the \textit{r} to \textit{z} bands and the FLAMINGOS \textit{J} band was used by GMC+2014 in their fit, an underestimation in opacity in optical wavelengths could result in the model-predicted eclipse depths implying a lower temperature than they should for high metallicity objects.
  However, as shown in Figure \ref{fig:figure3}, the differences produced by changes in metallicity  would likely not be large enough to reconcile our results.
  In addition, any differences in the \textit{J} band are likely to be even smaller \citep{Mann}. 
  No matter the changes we can make to theoretical stellar spectra, there is no single temperature that will match the reported depths in the \textit{TESS} and \textit{J} bands.
   
   Our preferred interpretation is that the result in GMC+2014 is a result of systematic errors.
   Systematic errors inherent to ground-based observation have been a problem when trying to infer temperature from precise eclipse measurements, most noticeably with hot Jupiters \citep{deMooij, Croll}.
   In \cite{hooton} the eclipse depth measured by one instrument is less than 50$\%$ of another for the eclipses of WASP-12\,b observed in the I-band.
   For our value of T$_{\rm eff,2}$, the predicted eclipse depth in the \textit{J} band is 0.0044, cf. a depth of 0.00737 reported by GMC+2014. This is a discrepancy of about 50\%, similar to the systematic error reported by \cite{hooton}.
   This suggests that systematic errors can produce the size of anomaly that we are finding.
   Going further, \cite{Hansen} conducted an analysis of eclipse depth uncertainties in regards to inferring atmospheric quantities and proposes an underestimation in error across all eclipse depth observations.
   Considering the need for precise measurements to properly constrain theoretical models, further observations by other ground-based and space-based instruments are needed to ensure accuracy.
   
\section{Summary}
   In this paper we have presented our analysis of the \textit{TESS} light curve of J0113+31 and derived orbital parameters by MCMC fitting.
   We do not confirm the hotter-than expected temperature reported by GMC+2014 for the M-dwarf companion.
   Our analysis found an effective temperature of ${\rm T}_{\rm eff,2}$ = 3208 $\pm$ 43K, a value that agrees well with those predicted by theoretical stellar models.
   Our preferred explanation for the discrepancy is that GMC+2014 under-estimated the systematic error in their ground-based measurement of the eclipse depth.

   Additional observations of J0113+31 among other EBLMs are planned using the recently-launched \textit{CHEOPS} satellite.
   The analysis of the high precision light curves observed by \textit{CHEOPS} of these objects will contribute towards the better understanding of low-mass stars using more accurate radii and temperatures.
   With its observational bandpass based in the visual part of the spectrum it would also be worthwhile to undertake further observation in the near-infrared to see if eclipse depths obtained in these different regimes still disagree, or if there are further possible causes for reported anomalous effective temperatures of low-mass stars.
   
\section*{Acknowledgements}
      MS and PM are supported by the UK Science and Technology Facilities Council (STFC) grant numbers ST/M001040/1 and ST/T506175/1.
      AHMJT has received funding from the European Research Council (grant agreement number 803193/BEBOP), the Leverhulme Trust (grant number RPG-2018-418), and from the STFC (grant number ST/S00193X/1).
      VKH is supported by a Birmingham Doctoral Scholarship, a studentship from Birmingham’s School of Physics $\&$ Astronomy and by a Fulbright Scholarship from the U.S. - Norway Fulbright Foundation. 
      We would like to thank the anonymous referee for their constructive and timely comments on the manuscript.

\section*{Data Availability}
      This paper includes data collected by the TESS mission, which is publicly available from the Mikulski Archive for Space Telescopes (MAST) at the Space Telescope Science Institure (STScI). 
      Funding for the TESS mission is provided by the NASA Explorer Program
    directorate. 
    STScI is operated by the Association of Universities for Research in Astronomy, Inc., under NASA contract NAS 5–26555.
    We acknowledge the use of public TESS Alert data from pipelines at the TESS Science Office and at the TESS Science Processing Operations Center.

%%%%%%%%%%%%%%%%%%%%%%%%%%%%%%%%%%%%%%%%%%%%%%%%%%



\bibliographystyle{mnras}
\bibliography{eblmvii} % if your bibtex file is called example.bib

%\begin{thebibliography}{99}
%\end{thebibliography}

%%%%%%%%%%%%%%%%%%%%%%%%%%%%%%%%%%%%%%%%%%%%%%%%%%

%%%%%%%%%%%%%%%%% APPENDICES %%%%%%%%%%%%%%%%%%%%%

%\appendix



%%%%%%%%%%%%%%%%%%%%%%%%%%%%%%%%%%%%%%%%%%%%%%%%%%

% Don't change these lines
\bsp	% typesetting comment
\label{lastpage}
\end{document}